\documentclass[12pt,a4paper]{article}
\usepackage{relsize}

\usepackage{amssymb}
\usepackage{amsmath,amsfonts}
\usepackage{graphicx}

\usepackage{pstricks}
\usepackage{floatflt}
\usepackage{cite}
\usepackage{wasysym}
\usepackage{hyperref}
\usepackage{graphics}
\usepackage{url}
\usepackage{amssymb}
\newcommand{\vareps}{\varepsilon}
\newcommand{\eqn}{equation}
\newcommand{\al}{\alpha}

\newcommand{\lb}{\left(}
\newcommand{\rb}{\right)}
\newcommand{\ph}{\hat{p}}
\newcommand{\D}{\mathcal{D}}
\newcommand{\M}{\mathcal{M}}

\newcommand{\nc}{\newcommand}

\nc{\beq}{\begin{equation}}
\nc{\eeq}{\end{equation}}
\nc{\bea}{\begin{eqnarray}}
\nc{\eea}{\end{eqnarray}}
\nc{\nn}{\nonumber}

\nc{\veps}{\varepsilon}
\nc{\eps}{\epsilon}
\nc{\as}{\alpha_s}
\nc{\cd}{\cdot}
\nc{\lag}{\cal L }
\nc{\matx}{\left|\cal {M}\right|^2}
\nc{\lqcd}{\Lambda_\textrm{QCD}}
\nc{\msbar}{\overline {\textrm{MS}}}
\nc{\really}{\stackrel{!}{=}}
\def\sla#1{\ifmmode%
\setbox0=\hbox{$#1$}%
\setbox1=\hbox to\wd0{\hss$/$\hss}\else%
\setbox0=\hbox{#1}%
\setbox1=\hbox to\wd0{\hss/\hss}\fi%
#1\hskip-\wd0\box1 }
\nc{\dsla}{\sla{\partial}}
\nc{\Dsla}{\sla{D}}

\newlength{\nseparation}
\begin{document}
\bibliographystyle{unsrt}
\thispagestyle{empty}
\def\thefootnote{\fnsymbol{footnote}}
\setcounter{footnote}{1}
\null
\vskip 0cm
\begin{center}

{\Large \boldmath{\bf
   Nagy-Soper subtraction: a review}
    \par} \vskip 2.5em {\large

{\sc Tania Robens}\\[1ex]
{\normalsize \it IKTP, TU Dresden, Zellescher Weg 19, 01069 Dresden, Germany}}
\par \vskip 2em
\end{center}\par

\noindent{\bf Abstract:}\\[0.25em]
\noindent 

In this document, we present a review on an alternative NLO subtraction scheme, based on the splitting kernels of an improved parton shower that promises to facilitate the inclusion of higher order corrections into Monte Carlo event generators. We give expressions for the scheme for massless emitters, and point to work on the extension for massive cases. As an example, we show results for the C parameter of the process $e^+\,e^-\,\rightarrow\,3$ jets at NLO which have recently been published as a verification of this scheme. We equally provide analytic expressions for integrated counterterms that have not been presented in previous work, and comment on the possibility of analytic approximations for the remaining numerical integrals.
\par
\null
\setcounter{page}{0}
\clearpage
\def\thefootnote{\arabic{footnote}}
\setcounter{footnote}{0}

\section{Introduction}	

With the successful start of the LHC, particle physics has entered an exciting era, where the recent discovery of a Higgs Boson\cite{Aad:2012tfa,Chatrchyan:2012ufa} is hopefully only one achieved milestone. An indispensable ingredient in LHC physics analyses is the understanding of Standard Model (SM) processes to high accuracy on a fully differential level. Appropriate tools for such predictions are Monte Carlo event generators including higher order contributions. The second ingredient towards a full next-to-leading order (NLO) description, i.e. the combination of parton level NLO calculations with parton showers, is equally well-understood and implemented in many current tools\cite{Nason:2012pr}.% \cite{Frixione:2002ik,Nason:2004rx,Frixione:2007vw}.

The implementation of NLO calculations into numerical tools exhibits a caveat stemming from the infrared divergence of real and virtual NLO contributions, which originate from different phase spaces: although in the sum of all contributions, the infinite parts exactly cancel, the behavior of the divergence needs to be parametrized, e.g. by infinitesimal regulators. The implementation of such regulators into numerical codes can result in large unphysical numerical uncertainties. A way to circumvent this problem is the introduction of subtraction schemes, which efficiently reshuffle the divergent terms such that a numerically stable evaluation becomes possible for both Born-type and real-emission kinematics contributions.

We here review such a subtraction scheme first presented in Refs. \cite{Chung:2010fx,Chung:2012rq}, which is based on the splitting kernels of an improved parton shower proposal\cite{Nagy:2007ty,Nagy:2008ns,Nagy:2008eq}, and therefore referred to as Nagy-Soper (NS) scheme in the following. We give the generic setup of the scheme as well as its differences with respect to other subtraction schemes\cite{Frixione:1995ms,Catani:1996vz}, and comment on possible further development and improvements. This manuscript is organized as follows: In Section \ref{sec:schemes}, we will briefly discuss important features in the setup of subtraction schemes. In Section \ref{sec:ns-sub}, we discuss the properties of our scheme, including possible further improvements. We report on the current status in Section \ref{sec:status} and summarize in Section \ref{sec:summ}.
\section{Subtraction Schemes}\label{sec:schemes}
Higher order subtraction schemes make use of factorization of the real-emission matrix element in the soft or collinear limits, where it can be decomposed according to\cite{Altarelli:1977zs,Bassetto:1984ik,Dokshitzer:1991wu}
\beq
\label{Factorizationm1toDim}
\left|{\cal M}_{m+1}(\hat p)\right|^2 \quad\longrightarrow\quad \D_\ell\,\otimes\,\left| {\cal M}_{m}( p) \right|^2,
\eeq
where $\D_\ell$ are the dipoles containing the respective singularity structure, and the symbol $\otimes$ denotes a correct convolution in color, spin, and flavor space. $\hat p/\, p$ denote momenta in $(m+1)/\,m$-parton phase space, respectively. One can then define subtracted contributions according to
\bea
\label{countertermfinite85}
\sigma^{\text{NLO}}&=&\underset{ \textrm {finite} }
{\underbrace{\int_{m+1}\left[
d\sigma^R-d\sigma^A\right]}}+\underset{ \textrm {finite} }
{\underbrace{\int_{m+1}\,d\sigma^A+\int_m\,d\sigma^V}}        %          \nn\\
%&=&\int_{m+1}\left[ d\sigma^R_{\vareps=0}-d\sigma^A_{\vareps=0}\right]+
%\int_{m}\left[\int_1\,d\sigma^A+d\sigma^V\right]_{\vareps=0},
\eea
where
\begin{alignat}{53}
\label{explicitexpressionsNLO}
\int_m\, \left[d\sigma^B\,+\,d\sigma^V\,+\,\int_1\,d\sigma^A\right]& =\int    dPS_m
\left[\left| {\cal M}_{m} \right|^2\,+\,\left| {\cal M}_{m} \right|^2_{\textrm{one-loop}}\,+\,
\sum_\ell\,\mathcal{V}_\ell\,\otimes\,\left| {\cal M}_{m} \right|^2\right], \notag \\
\int_{m+1}\,\left[ d\sigma^R - d\sigma^A \right]&=\int dPS_{m+1} \left[\left| {\cal M}_{m+1} \right|^2 \,-\, \sum_\ell\,D_\ell\,\otimes\,\left| {\cal M}_{m} \right|^2\right],  
\end{alignat}
and where $\int\,d\,PS$ indicates the integration over the respective phase space, including all symmetry and flux factors.
Here, $d\sigma^B,\,d\sigma^V,\,d\sigma^R$ denote the Born-type, virtual contribution and real-emission parts of the calculation, while real-emission subtraction terms are summarized as $d\sigma^A$. Furthermore, 
as $\left|{\cal M}_{m+1}\right|^2$ and 
$\left| {\cal M}_{m} \right|^2$ live in different
phase spaces, a mapping of their momenta via a mapping function needs to be introduced. The subtraction term
$\D_\ell$ and its one-parton integrated counterpart $\mathcal{V}_\ell$ are related by 
\beq
\mathcal{V}_\ell\,=\, \int\,d\xi_p \,\D_{\ell},
\eeq
where $d\xi_p$ is an unresolved one-parton integration measure. The following ingredients are therefore needed in order to define a subtraction scheme:
%\begin{itemize}
%\item 
(a) a suitable mapping from $(m+1)$ to $m$ parton phase space which guarantees energy-momentum conservation as well as on-shellness, and
%\item 
(b) an efficient parametrization of the one-parton integration measure $d\xi_p$.
%\end{itemize}
While  the number of reevaluations of the underlying Born matrix element in the real emission subtractions in Eqn. (\ref{explicitexpressionsNLO}) is determined by (a), the complexity of the integrated counterterms depends on (b). Currently, two major schemes for next-to-leading order subtraction are on the market, namely
%\begin{itemize}
%\item 
the Catani-Seymour (CS) dipole scheme, first proposed in Ref. \cite{Catani:1996vz} (expressions for massive emitters have been published in Ref. \cite{Catani:2002hc}), and
%\item 
the Frixione-Kunszt-Signer (FKS) subtraction proposed in Ref. \cite{Frixione:1995ms}.
%\end{itemize}
%In the last decade, both formalisms have been automized, either as stand-alone packages or within multi-purpose Monte Carlo event generators including higher order corrections; we here only briefly list the major differences with respect to the scheme proposed here, and refer the reader to the literature for more details. 

Our scheme is characterized by the following ingredients:
\begin{enumerate}
\item \label{it:kernel} we use the splitting kernels of an improved parton shower as a basis for the real emission subtraction terms, which promises to facilitate the matching to the improved parton shower,
\item \label{it:mapping} we apply a momentum mapping which leads to an overal scaling behavior $\sim\,N^2$ for a process with $N$ partons in the final state.
\end{enumerate}

The number of matrix element reevaluations is reduced by a factor proportional to the number of final state particles of the process with respect to the CS scheme.
%\item 
While a similar scaling behavior is inherent in the FKS subtraction scheme, we do not need to reparametrize the phase space for each emitter/emitted parton pair.
%\end{itemize}
Unfortunately, these advantages of our scheme come with more complex expressions for the integrated counterterms. We will comment on each of these features seperately in more detail below.

\section{Nagy-Soper subtraction: Setup and relation to improved parton shower}\label{sec:ns-sub}
In this section, we briefly recapitulate the setup of our scheme, with a special focus on the mapping leading to the improved scaling behavior, as well as further treatment of the integrated counterterms and the matching to the parton shower. We use the following conventions:
\begin{itemize}
\item four-momenta in the Born-type kinematics are denoted by unhatted quantities $p_i$, while the real emission phase space momenta are denoted by hatted quantities $\ph_i$;
\item initial state momenta are denoted $p_a$ and $p_b$, where $Q\,=\,p_a+p_b$ and with $Q^2$ being the center-of-mass energy, with equivalent relations in the real emission phase space;
%\item for radiation off initial state particles, we always take $p_a$ as the mother parton;
\item generally, $\ph_\ell$ labels the emitter, with $\ph_j$ being the emitted parton and $\ph_k$ the spectator. Note that we only need to specify spectators in the case of interference term singularities (c.f. the discussion below). 
\end{itemize}

\subsection{Scheme setup}
The scheme discussed here uses the splitting kernels of an improved parton shower\cite{Nagy:2007ty,Nagy:2008ns,Nagy:2008eq} as a basis for the subtraction terms, as they exhibit the same behavior in the singular limits. We can therefore write
\beq
\label{QCDFactorizationm1tVm}
\mid {\cal M}_\ell(\{\hat p, \hat f\}_{m+1})\rangle\,=\,t^\dagger_\ell(f_\ell \to \hat f_\ell + 
\hat f_{j})\,V^\dagger_\ell(\{\hat p, \hat f\}_{m+1})\,\mid {\cal M}(\{ p,  f\}_{m})\rangle,
\eeq
where we adopt the matrix-element notation introduced in Ref. \cite{Nagy:2007ty}. Here,\\  $\mid {\cal M}_\ell(\{\hat p, \hat f\}_{m+1})\rangle$ and $\mid{\cal M}(\{ p,  f\}_{m})\rangle$ denote the matrix elements in real emission ($m+1$) and Born-type ($m$) phase space and $V_\ell,\,t_\ell$ the factorization operators in color and spin space, respectively (explicit forms of the $V_\ell$'s can be obtained from Ref. \cite{Nagy:2007ty}). We first discuss the treatment of the splitting functions in spin space. For fermionic emitters, the splitting functions are diagonal in helicity space; therefore, the real emission subtraction terms are directly given by the spin averaged functions
\begin{\eqn*}
{\textstyle \overline{W}_{\ell\,\ell}\,=\,\frac{1}{2}v^2_\ell,}
\end{\eqn*}
with $v_\ell^2$ being defined through the relation
\begin{alignat}{53}
%\boxed{
{\textstyle
\langle \{\hat s\}_{m+1}\mid V^\dagger_\ell(\{\hat p, \hat f\}_{m+1})\mid \{s\}_m\rangle\,=\,
\left(\prod_{n\notin\{\ell,j=m+1\}} \delta_{\hat s_n,s_n}\right)\,
v_\ell (\{\hat p, \hat f\}_{m+1},\hat s_{j},\hat s_{\ell},s_\ell).
}
%}
\label{eq:vel_def}
\end{alignat}
In case of gluonic emitters, the situation is more complicated, as information of the gluon polarization needs to be retained. In addition, if the emitted particle is a gluon, we encounter soft/collinear divergences from interference terms.
%, as can be seen from figure \ref{amplitudesquare_Softdiagram_qqg}. 
%\input{fig1}
In our scheme, these are treated using dipole partitioning functions $A_{\ell k}$, which redistribute the singularities to contributions $W^{(\ell)}_{\ell k},\,W^{(k)}_{\ell k}$, where $p_\ell,\,p_k$ take over the kinematic role of the mother parton in the mapping, respectively (this procedure is discussed in detail in Ref. \cite{Nagy:2008eq} within the framework of the parton shower). The subtraction term is then split into a purely collinear and a soft/collinear part, and we have
\beq
\label{splitsplittingfunctions335}
\langle \nu'| W_{\ell\ell}- W_{\ell k}|\nu\rangle \,=\,\langle \nu'| \left( {W}_{\ell\ell} - {W}_{\ell\ell}^{\textrm {eik}} \right)
+ \left({W}_{\ell\ell}^{\textrm {eik}}  - {W}_{\ell k}\right)|\nu\rangle,
\eeq
where $|\nu\rangle,\,|\nu'\rangle$ denote the gluon polarization of the mother parton connected to the Born-type matrix element in Eqn. (\ref{QCDFactorizationm1tVm}). $W^\text{eik}_{\ell \ell},\,W_{\ell k}$ are related to the eikonal splitting functions $v^\text{eik}_\ell\,\sim\,\frac{\vareps(\ph_j,\hat{Q})\cdot\,\ph_\ell}{\ph_\ell\,\cdot\,\ph_j}$ via
\begin{\eqn*}
W^\text{eik}_{\ell \ell}\,\sim\,|v_\ell^\text{eik}|^2,\,W_{\ell\,k}\,\sim\,A_{\ell k}\,v_\ell^\text{eik}\,\lb v_k^\text{eik} \rb^*.
\end{\eqn*}
  While the first part of the right-hand side of Eqn.~(\ref{splitsplittingfunctions335}) contains the collinear singularity, where the polarization information needs to be retained, the second part corresponds to the soft and soft/collinear term, where we can again average over initial state spins in the subtraction term. With a specific choice of dipole partitioning functions, the latter is given by
\beq
\label{interferencespinaveragedsplittingfunction}
%\boxed{
\Delta W_{\ell k} \,=\, \overline{W}_{\ell\ell}^{\textrm {eik}}  - \overline{W}_{\ell k} \,=\,4\,\pi\,\al_s
\frac{2\, \hat p_\ell\cdot\hat p_k\, \hat p_\ell\cdot\hat Q } 
{\hat p_\ell\cdot\hat p_j\,
\left(\hat p_j\cdot\hat p_k\, \hat p_\ell\cdot\hat Q+\hat p_\ell\cdot\hat p_j\,\hat p_k\cdot\hat Q \right)}
%}
\eeq
where the spin-averaged eikonal factor has the explicit form
\beq
\label{spinaveragedsplittingfunctions_Wll_eikonal}
{\textstyle
\overline{W}_{\ell\ell}^{\textrm {eik}}\,=\, 4\,\pi\,\as\, 
\frac{\hat p_\ell\cd  D(\hat p_{j},\hat Q) \cd  \hat p_\ell}
     { (\hat p_{j}\cd \hat p_\ell)^2 }
}
\eeq
with the transverse projection tensor 
$
D^{\mu\nu}(\hat p_j,\hat Q) \,=\,- g^{\mu\nu} + \frac{\hat p_j^\mu\, \hat Q^\nu + \hat Q^\mu\, \hat p_j^\nu}
{\hat p_j\cd \hat Q}- \frac{\hat Q^2\, \hat p_j^\mu\, \hat p_j^\nu}{(\hat p_j\cd \hat Q)^2}$. 
The color factors are 
\begin{eqnarray*}
C(\hat{f}_{\ell},\hat{f}_{j})\,=\,
\begin{cases}
C_{F}& (\hat{f}_{\ell},\hat{f}_{j})\,=\,(q,g), (g,q),\\
C_{A}& (\hat{f}_{\ell},\hat{f}_{j})\,=\,(g,g),\\
T_{R}& (\hat{f}_{\ell},\hat{f}_{j})\,=\,(q,\bar{q}),\\
\end{cases}
\end{eqnarray*}
for the collinear and
\begin{\eqn}\label{eq:clk}
{\textstyle
C_{\ell\,k}\,\equiv\,-\frac{1}{2}\,\left[t^{\dagger}_{k}\,\otimes\,t_{\ell}+ t^{\dagger}_{\ell}\,\otimes\,t_{k}\right]}
\end{\eqn}
for the interference terms respectively. Within the framework of Monte Carlo event generators, which usually contain the physics information on the amplitude level, the real emission subtraction terms can equally be implemented on such a level. A detailed description on how to implement the dipole partitioning functions in this case can e.g. be found in Ref. \cite{Nagy:2008eq}. The expressions presented here for the real emission subtraction terms are therefore more tailored for codes working on the squared matrix element level.

\subsection{Momentum mapping}
The specific mapping between the real emission and Born-type kinematic phase spaces for final state emitters is the major cause of the improved scaling behavior. We briefly review it here; for a more detailed discussion, including initial state mappings as well as the relation to the mapping in the corresponding parton shower, we refer the reader to Refs. \cite{Chung:2010fx,Chung:2012rq}.

For the final state mappings, we have to consider two different cases:
\begin{\eqn*}
Q^2\,=\,2\,P_\ell\cdot\,Q-P_\ell^2,\;\hspace{10mm} Q^2\,>\,2\,P_\ell\cdot\,Q-P_\ell^2,
\end{\eqn*}
which we distinguish by the parameter $a_\ell\lb P_\ell,Q\rb\,=\,\frac{Q^2}{2\, P_\ell\,\cdot\,Q-P_\ell^2}$, where $P_\ell\,=\,\ph_\ell+\ph_j$. 

For $a_\ell\,=\,1$, we have
\begin{\eqn}\label{eq:fin_map_a1}
p_{\ell}\,=\,\frac{1}{1-y_\ell}\,\left(\ph_{\ell}+\ph_{j}-y_\ell\,Q\right),\;p_n\,=\,\frac{\ph_{n}}{1-y_\ell}\, 
\end{\eqn}
with $n \neq\,(\ell,j)$. 

For $a_\ell\,\neq\,1$, we map the momenta according to
\begin{eqnarray}
\label{eq:fin_map}
p_\ell&=&\frac{1}{\lambda_\ell}\,(\hat p_\ell + \hat p_j)-\frac{1 - \lambda_\ell + y_\ell}{2\, \lambda_\ell\, a_\ell}\, Q,\nn\\
p_n^\mu&=&\Lambda (K,\hat{K})^\mu{}_\nu \,\hat p^{\nu}_{n} ,\quad n\notin\{\ell,j=m+1\}
\end{eqnarray}
with
\begin{eqnarray}\label{eq:LTini}
&&{\textstyle \Lambda(K,\hat K)^{\mu}_{\;\;\nu} \,=\,g^{\mu}_{\;\;\nu}\,-\,\frac{2\,( K+\hat K)^{\mu}\,(K+\hat K)_{\nu}}{(K+\hat K)^{2}}\,
+\,\frac{2\,{K}^{\mu}\,\hat{K}_{\nu}}{\hat K^{2}}\,}.
\end{eqnarray}
The parameter $y_\ell$, which serves as a virtuality measure of the splitting, is given by 
\begin{\eqn*}
{\textstyle
y_\ell = \frac{P_\ell^2}{2\, P_\ell\cdot Q - P_\ell^2}.}
\end{\eqn*}
We furthermore introduced
%\begin{eqnarray}\label{eq:ydef}
%&&
${\textstyle \lambda_\ell\,=\,\sqrt{\left(1+y_\ell\right)^2-4\,a_\ell\,y_\ell},}\;{\textstyle K\,=\, Q - p_\ell,\;\hat{K}\,=\, Q - P_\ell.}$
%\end{eqnarray}
Note that it is the {\sl global} mapping for all remaining particles in Eqn. (\ref{eq:fin_map}) that is responsible for the reduced number of Born-type matrix reevaluations needed in our scheme. Furthermore, it is worth to notice that the mappings for $a_\ell\,=\,1,\;a_\ell\,\neq\,1$ correspond to two distinct prescriptions which do not coincide in the limit $a_\ell\,\rightarrow\,1$.

%Note that the global mapping introduced in eqn \ref{} is the main reason for the

\subsection{Final expression and scaling behavior}

In this subsection, we present the expressions needed for the application of our scheme, and discuss the resulting scaling behavior for momentum mappings in the real emission subtraction terms, as well as possible further improvements of the scaling. We start with the expression for a total cross section/observable
\begin{alignat}{53}
\label{eq:sig_nlo}
\sigma_{ab}^{\text{NLO}}(p_a,p_b,\mu_F^2)&= \int_{m+1} \left[
  d\sigma_{ab}^{R}(\ph_a,\ph_b) -
  d\sigma_{ab}^{A}(\ph_a,\ph_b) \right]  \,F_J^{(m+1)}(\ph_{m+1})                  \notag \\
&\hspace{-12mm}+ \int_{m}\, \left[ d\sigma^{V}_{ab}(p_a,p_b)+
  \int_1d\sigma^{A}_{ab}(\ph_a,\ph_b)+d\sigma_{ab}^{C}(p_a,p_b,\mu_F^2)\right]_{\vareps=0}\,F_J^{(m)}(p_m),
\end{alignat}
where $\int_1 d\sigma^A_{ab}+ d\sigma^C_{ab}$ can be written as
\begin{alignat}{53}\label{IKP2842010}
&\hspace*{-10mm}\int_{m}  \left[ \int_{1} d\sigma^A_{ab}(\ph_a,\ph_b)+ d\sigma^C_{ab}(p_a,p_b,\mu_F^2) \right]    \notag \\
=&
\int_m  d\sigma_{ab}^{B}(p_a,p_b) \otimes { I}(\vareps)
+ \int_0^1 dx \int_m  d\sigma_{ab}^{B}(x\,\ph_a,p_b)\otimes
\left[ { K}^a(x\,\ph_a) +   { P}(x,\mu_F^2)  \right]                             \notag \\
+&  \int_0^1 dx \int_m d\sigma_{ab}^{B}(\ph_a,x\,\ph_b)\otimes
\left[ { K}^b(x\,\ph_b) + { P}(x,\mu_F^2)  \right],
\end{alignat}
and the jet functions $F_J^{(m+1)}(\ph_{m+1}),\,F_J^{(m)}(p_{m})$ exhibit standard properties in the soft or soft/collinear limits, i.e. $F_J^{(m+1)}(\ph_{m+1})\,\rightarrow\,F_J^{(m)}(p_{m})$.
The above equation defines the insertion operators $I(\vareps),\,K(x),\,
P(x;\mu_{F})$ on a cross section level, and further combinatorial factors $N_{m},N_{m+1}$ are related to the integrated subtraction terms $\mathcal{V}$ via
\begin{\eqn}\label{eq:v_ikp}
{\textstyle
\sum\,\mathcal{V}\,=\,\frac{1}{x}\,\frac{N_{m}}{N_{m+1}}\lb
I\,+\,K\,+\,P \rb.}
\end{\eqn}
For the real emission subtraction terms, we have
\begin{\eqn}\label{eq:master_sub}
d\sigma^{A}_{ab}(\ph_a,\ph_b)\,=\,d\sigma^{A,a}_{ab}(\ph_a,\ph_b)+d\sigma^{A,b}_{ab}(\ph_a,\ph_b)
+\sum_{\ell\,\neq\,a,\,b} d\sigma^{A,\ell}_{ab}(\ph_a,\ph_b),
\end{\eqn}
with the sum over all possible final state emitters $\ph_\ell$ (the expressions for initial state emitters can readily be obtained from Ref. \cite{Chung:2010fx} and will not be repeated here). We then have
\begin{eqnarray}\label{eq:counter_fin}
d\sigma^{A,\ell}_{ab}(\hat{p}_a,\hat{p}_b)&=&\frac{N_{m+1}}{\Phi_{m+1}}
\sum_{j}\Bigg\{
\left[\D_{gqq}(\ph_{j})\delta_{g;q,q_{j}}+\D_{ggg}(\ph_{j})
\delta_{g;g,g_{j}}\,\right]|\M_{\text{Born},g}|^{2}(p_{a},p_{b};p_{n})\nn\\
&&\,+\left[ \D_{qqg}(\ph_{j})\delta_{q;g,q_{j}}\,+\,\D_{qqg}(\ph_{j})
\delta_{q;q,g_{j}}(\ph_{j})\right]|\M_{\text{Born},q}|^{2}(p_{a},p_{b};p_{n})\Bigg\},\nn\\
&&
\end{eqnarray}
where $\Phi_{m+1}$ denotes the flux factor. The $\delta_{f_\ell;\hat{f}_\ell\,\hat{f}_j}$ functions ensure the existence of the respective splittings in flavor space for the splitting $f_\ell\,\rightarrow\,\hat{f}_\ell\,\hat{f}_j$. For a given flavor $\hat{f}_\ell$, obviously only terms $\sim\,\mathcal{D}_{f_\ell;\hat{f}_\ell\,\hat{f}_k}$  contribute.
 Following the previous discussion on the treatment of singularities stemming from soft or soft/collinear interference, the subtraction terms can be split into collinear and interference parts:
 \begin{\eqn}\label{eq:delljtot}
\D_{f_\ell \hat{f}_\ell \hat{f}_{j}}(\ph_\ell, \ph_j)\,=\,\D^{\text{coll}}_{f_\ell \hat{f}_\ell \hat{f}_{j}}(\ph_\ell, \ph_j)\,+\,\delta_{\hat{f}_{j},g}\sum_{k\,\neq\,(\ell,j)}\D^{\text{if}}(\ph_\ell,\ph_j,\ph_{k}),
\end{\eqn}
where $\D^{\text{if}}(\ph_\ell,\ph_j,\ph_{k})$ denotes an interference contribution with $\ph_{k}$ acting as a spectator. The improved scaling behavior in our scheme now stems from the fact that each of the contributions in Eqn. (\ref{eq:delljtot}) requires exactly {\sl one} global mapping, i.e. the number of mappings and thereby matrix reevaluations in the real emission subtraction terms behaves like $\sim\,\#(\ell j)$. Table \ref{tab:comb} explicitely lists the number of reevaluations/ mappings needed for a given number of partons in the final state in the CS and NS schemes.
\begin{table}
%\beq
\begin{tabular}{c||c c||c}
\textrm{\small emitter, spectator}&{ \textrm{\bf CS},($\ell\,j$)}& {  \textrm{\bf CS},$k$}&{  \textrm{\bf NS},($\ell\,j$)}\\ \hline\hline
\textrm{fin,fin}&$\binom{N'}{2}$&$(N'-2)$&$\binom{N'}{2}$\\
\textrm{fin,ini}&$\binom{N'}{2}$&$2$&$-$ \\
\textrm{ini,fin}&$2\,N'$&$(N'-1)$&$2\,N'$\\
\textrm{ini,ini}&$2N'$&$1$&$-$ \\ \hline
\textrm{total}&$N'^{2}(N'+3)/2\,=\,$&&$N'(N'+3)/2\,=\,$\\
${\scriptstyle (\sum_\text{comb's} (\ell j)\times (k))}$&$({ N}+1)^2({  N}+4)/2$&&$({ N}+1)({ N}+4)/2$\\ \hline
$\sim$&${  N^3}/2$&&${  N^2}/2$
\end{tabular}\label{tab:comb}
%\eeq
\caption{Combinatorial counting of Born-type reevaluations in the CS and NS schemes, with $N'$ being the number of real emission and $N=N'-1$ the number of Born-type final state particles. The expressions are exact for processes where all partons are gluons, while processes with quarks generally require less mappings.}
\end{table}

While this already sufficiently demonstrates the difference of matrix-element reevaluations needed in these two schemes, this scaling behavior can in principle be improved even further. In Ref. \cite{Frederix:2009yq}, the authors show that within the MadFKS environment, a constant scaling behavior can be achieved; i.e., for certain types of processes, the number of reevaluations of underlying Born-type matrix elements in the real-emission subtraction terms remains constant. In Ref. \cite{Chung:2012rq}, we argue that within our scheme the same constant scaling behavior is possible. In both cases, this follows from the option to partition the real-emission phase space according to Eqn. (6.7) in Ref. \cite{Frederix:2009yq}
\begin{\eqn}\label{eq:fks_sym}
{\textstyle
d\sigma^{(n+1)}(r)\,=\,\sum_{(i,j)\in\overline{\mathcal{P}}_\text{FKS}}\xi_{ij}^{(n+1)}(r)d\sigma_{ij}^{(n+1)}(r),}
\end{\eqn}
followed by an evaluation of only non-redundant contributions in the real emission part of the process, here labelled by $\overline{\mathcal{P}}_\text{FKS}$. Our choice of dipole partitioning function leads to a unique mapping per non-redundant contribution, therefore in principle allowing for the same constant scaling behavior. A direct implementation of this into a numerical code is in the line of future work.
\subsection{Subtraction terms and integrated counterterms}
Explicit expressions for the subtraction terms $\D_{\ell jk}$ as well as the integrated counterterms $\mathcal{V}_{\ell k}$, including the definition of all variables,  are readily available from Refs. \cite{Chung:2010fx,Chung:2012rq} and will not be repeated here. Instead, we devote this subsection to a more detailed discussion of the leftover finite parts in the integrated counterterms that are currently evaluated numerically. The existance of finite remainders in these terms is a direct consequence of the modified mapping which leads to the improved scaling behavior discussed above. Although this constitutes a slight modification with respect to standard schemes such as CS and FKS, it poses no impediment for the implementation of our scheme. %However, it still constitutes a slight modification from the implementation of our scheme with respect to standard schemes such as CS and FKS. 

As an example, we list the expression for the final-final state interference term (Eqn. (62) in Ref. \cite{Chung:2012rq})
\begin{eqnarray}\label{eq:vif}
\mathcal{V}_{\ell k}^{\text{if}}&=&\mu^{2\,\vareps}\,C_{\ell k}\,\int\,d\xi_{p}\,(\Delta\,W_{\ell k})\,=\,
\lb\frac{2\,\mu^{2}\,\pi}{p_{\ell}\cdot Q}\rb^{\vareps}\frac{\al_{s}}{\pi}\,\frac{1}{\Gamma(1-\vareps)}C_{\ell k}\nn\\
&&\times \left\{\frac{1}{2\,\vareps^{2}}\,+\,\frac{1}{\vareps}\,\left[1\,+\,\frac{1}{2}\,\ln\lb \tilde{a}^{(\ell k)}_{0}+a_\ell\rb\right]\,\right.\left. -\,\frac{\pi^{2}}{6}\,+\,3\,-\,2\,\ln\,2\,\ln\lb \tilde{a}^{(\ell k)}_{0}+a_\ell \rb\right. \nn \\
&&\left.\,+\,\frac{1}{\pi}\left[I^{(b)}_\text{fin}\lb\frac{\tilde{a}^{(\ell k)}_{0}}{a_\ell}\rb\,+\,I^{(d)}_\text{fin}(a_\ell,\tilde{a}^{(\ell k)})\,+\,I^{(e)}_\text{fin}(a_\ell)\right]\right.\nn\\
&&\left. +\,\ln\,a_\ell\,\left[2\,\ln\,2\,-\,\frac{1}{4}\,\ln\,a_\ell\,+\,\frac{1}{2}\,\ln\,\lb \tilde{a}^{(\ell k)}_{0}\,+\,a_\ell\rb \,+\,1\right]\,\right\},\nn\\
&&
\end{eqnarray}
where the integrals $I^{(b)}_\text{fin}(b),\,I^{(d)}_\text{fin}(a_\ell,\tilde{a}^{(\ell,k)}),\,I^{(e)}_\text{fin}(a_\ell)$ (c.f. Eqn. (63) in Ref. \cite{Chung:2012rq}) are up to two-dimensional integrals depending on Lorentz-invariant products of the four-momenta in real emission phase space. 
Similar integrals, which depend on maximally three additional integration variables, occur in the case of initial-initial state interference terms. These additional integrations, which can readily be implemented into numerical codes, still denote a non-standard modification with respect to standard subtraction schemes; therefore, we briefly comment on possible approximations of these integrals through analytic functions or grids.
\subsubsection*{Analytical or numerical approximations of leftover finite integrated counterterms}

To start, we want to comment that two of the integrals which have been presented previously are now (partially) available analytically\footnote{We thank S. Dittmaier for useful comments regarding this.}. For the initial-final state interference term (Eqn. (111) in Ref. \cite{Chung:2010fx}), we have
\begin{eqnarray*}
\lefteqn{I_\text{fin}(x,\tilde{z})\,=\,\pi\,\delta(1-x)\,\times}\nonumber\\
&&\Bigg\{-\frac{\pi^2}{12}-\frac{1}{2}\left[\log^2\lb\frac{1-\sqrt{1-\tilde{z}_0}}{\tilde{z}_0} \rb-\frac{1}{2}\log^2\lb\frac{1+\sqrt{1-\tilde{z}_0}}{1-\sqrt{1-\tilde{z}_0}}  \rb  \right]\nonumber\\
&&+ \text{Li}_{2}\lb -\sqrt{1-\tilde{z}_0} \rb\,-\,\text{Li}_{2}\lb-\frac{1-\tilde{z}_0+\sqrt{1-\tilde{z}_0}}{\tilde{z}_0}  \rb -2\,\ln\,2\,\ln \tilde{z}_0 \Bigg\}\nonumber\\
&&+\,\frac{1}{(1-x)_{+}}\,\int^{1}_{0}\,\frac{dy'}{y'}\,\left\{\left[\int^{1}_{0}\,\frac{dv}{\sqrt{v\,(1-v)}}\,\frac{\tilde{z}}{N(x,y',\tilde{z},v)}\right]-\pi\right\},
\end{eqnarray*}
where we have used results from Ref. \cite{Beenakker:1988bq} to obtain the above expression. Along similar lines, the expression for  $I^{(b)}_\text{fin}(b)$ simplifies to
\begin{eqnarray*}
\lefteqn{I^{(b)}_\text{fin}(b)=-\frac{\pi^3}{12}\,+\,\pi\Bigg\{ 2\,\ln\,2\,\ln\,(1+b)}\nn\\
&&+ \frac{1}{2}\left[2 \text{Li}_2\left(-\sqrt{\frac{b}{b+1}}\right)-2 \text{Li}_2\left[-\sqrt{b} \left(\sqrt{b}+\sqrt{b+1}\right)\right]\right.\nonumber\\
&&\left.+\log ^2\left(\sqrt{b+1}
   \left(\sqrt{b+1}-\sqrt{b}\right)\right)-\frac{1}{2} \log ^2\left(\frac{\sqrt{b}+\sqrt{b+1}}{\sqrt{b+1}-\sqrt{b}}\right)\right] \Bigg\}.
\end{eqnarray*}
Both expressions have been implemented in the numerical codes that were used to obtain the results presented in Refs. \cite{Chung:2010fx,Chung:2012rq}, and we have explicitely verified that using these, we reproduce the published results within per mil accuracy.

We are then left with in total six integrals that call for numerical evaluation:
\begin{itemize}
\item 
%\begin{\eqn*}
$I_3(a_\ell),I_\text{fin}(a_\ell)$
%\end{\eqn*}
from the integration of the collinear $qqg,\,ggg$ splitting functions in the final state
\item
%\begin{\eqn*}
$I_\text{fin}(\tilde{z},x)$
%\end{\eqn*}
from the integration of the initial-initial interference term
\item
%\begin{\eqn*}
$I_\text{fin}^{(c)}(\tilde{a}),\,I_\text{fin}^{(e)}(a_\ell),\,I^{(d)}_\text{fin}\,(a_\ell,\tilde{a})$
%\end{\eqn*}
from integration of the final-final interference term.
\end{itemize}
We managed to obtain analytic approximations for all integrals which depend on a single input variable by polynomials; the fit functions reproduce the numerical results within $\mathcal{O}(10^{-6})$ accuracy, and we have verified their correctness by reproducing the results in Refs. \cite{Chung:2010fx,Chung:2012rq}. We have equally approximated one of the remaining integrals $I_\text{fin}(\tilde{z},x),\,I^{(d)}_\text{fin}\,(a_\ell,\tilde{a})$ via interpolating grids. We therefore expect to present a complete set of analytic approximations for all integrals depending on a single variable as well as grid approximations for the other integrals within the near future\cite{markusandme}. However, we want to emphasize that the evaluation of the real emission contributions is significantly more computationally expensive than the evaluation of the Born-type kinematics contribution, even if the additional finite counterterms are evaluated numerically.

\subsection{Combination with the shower}

In this subsection, we want to comment on an additional possible advantage of our scheme, i.e. the mapping of a NLO calculation at parton level with the improved parton shower suggested in Refs. \cite{Nagy:2007ty,Nagy:2008ns,Nagy:2008eq}. We will briefly review the reasoning why this should indeed pose a possible improvement with respect to matching with a shower that uses different splitting kernels. In a standard parton shower, the further splittings of the partons which participate in the hard interaction are typically described on a classical level, with interferences only taken into account approximately via angular ordering, while spin and subleading color contributions are neglected. In Ref. \cite{Nagy:2007ty}, a new shower algorithm has therefore been proposed which promises to include all these subleading contributions and equally describes the splitting on a full quantum-mechanical level. Agreement of this shower with classical descriptions in the classical limit has been proven in Ref. \cite{Nagy:2008ns}. Although an implementation of this shower is underway\cite{zoltandave}, it has not yet made its way into one of the publically available Monte Carlo event generators\footnote{We also want to mention that further improvement regarding the inclusion of subleading color has been presented e.g. within the Herwig++ framework\cite{Platzer:2012np}.}. Nevertheless, we here briefly review the advantages that are incorporated in matching a parton-level NLO calculation employing a specific subtraction scheme with a parton shower using exactly the same splitting functions. In this, we largely follow the arguments and notations presented in Ref. \cite{Hoeche:2011fd}.

For the inclusion of a suitable matching between NLO corrected hard processes and a parton shower, one can symbolically write for the total NLO cross section
\begin{eqnarray}\label{eq:nlo_match}
\lefteqn{\sigma_\text{NLO}\,=\,}\nn \\
&&\int\,d\Phi_B\overline{B}\lb \Phi_B \rb\left[ \overline{\Delta}(t_0,\mu^2_F)\,+\,\int^{\mu_F^2}_{t_0}d\Phi_1\,\frac{D(\Phi_B,\,\Phi_1)}{B(\Phi_B)}\overline{\Delta}(t,\mu^2_F)  \right]\nn\\
&&\,+\,\int\,d\Phi_R\,\left[ R(\Phi_R)-D(\Phi_B,\Phi_1) \right],
\end{eqnarray}
where
\begin{eqnarray*}
\overline{B}(\Phi_B)&=&B(\Phi_B)+V(\Phi_B)+I^{(S)}(\Phi_B)\,+\,\int\,d\Phi_1\,\left[ D(\Phi_B,\Phi_1)-D^{(S)}(\Phi_B,\Phi_1) \right],\\
\overline{\Delta}(t,t')&=&\exp\,\left[ -\int^{t'}_t\,d\Phi_1\,\frac{D(\Phi_B,\Phi_1)}{B(\Phi_B)}\right].
\end{eqnarray*}
Here, $B(\Phi_B),\,V(\Phi_B),\,I^{(S)}(\Phi_B)$ denote the Born, virtual, and integrated counter-term contribution to the parton-level NLO cross section, while $R(\Phi_R)$  and $D^{(S)}(\Phi_B,\Phi_1)$ are the real emission contribution and subtraction terms. $\Phi_B,\,\Phi_R,\,\Phi_1$ refer to the Born, real emission and one-particle phase space, respectively. The correct matching to the parton shower then follows from the use of the modified Sudhakov form factor $\overline{\Delta}(t,t')$, where the convolution in Eqn. (\ref{eq:nlo_match}) guarantees a correct NLO description throughout the whole real emission phase space. The differences between the Powheg\cite{Nason:2004rx,Frixione:2007vw} and MC@NLO\cite{Frixione:2002ik} methods as discussed in Ref. \cite{Hoeche:2011fd} then basically stem from the choice of $D(\Phi_B,\Phi_1)$, which determines the weight in the matrix-element correction procedure. We here do not want to comment on differences or (dis)advantages of any of the above schemes, but rather point to the fact that choosing $D(\Phi_B\,\Phi_1)\,=\,D^{(S)}\,(\Phi_B,\Phi_1)$, the additional integration in the modified Born contribution $\overline{B}(\Phi_B,\,\Phi_1)$ becomes trivial. This is the choice labelled "MC@NLO" in Ref. \cite{Hoeche:2011fd}, and details on the implementation for the case of CS subtraction terms can be found in Refs. \cite{Hoeche:2011fd,Hoeche:2012fm}. We currently see this as the major advantage of using the same splitting kernels in the subtraction scheme and the shower\footnote{In Ref. \cite{Hoeche:2011fd}, the authors additionally make the argument that this choice is indeed needed in order to correctly describe infrared-divergent subleading color contributions.}.

\section{Current status}\label{sec:status}
\subsection{Massless emitters}
The massless scheme including all real emission and integrated counterterms has been presented in Refs. \cite{Chung:2010fx,Chung:2012rq}, where we equally give detailed calculations for several standard processes. Explicit NLO QCD results are provided for
%\begin{center}
%\begin{itemize}
%\item 
single W production,
%\item 
$e^{+}\,e^{-}\,\rightarrow\,(2,3)$ jets,
%\item 
$H\,\rightarrow\,g\,g$ and $g\,g\,\rightarrow\,H$, and 
%\item 
$e\,q\,\rightarrow\,e\,q$.
%\end{itemize}
%\end{center}
%\subsubsection*{Example: $e^+\,e^-\,\rightarrow\,3$ jets}
 As an example, we here show the results for the C parameter in the process $e^+\,e^-\,\rightarrow\,\text{3 jets}$, which have originally been presented in Ref. \cite{Chung:2012rq}.
%\begin{\eqn*}
%$e^+\,e^-\,\rightarrow\,\text{3 jets}$.
%\end{\eqn*}
For this, we have the leading order contribution
\begin{\eqn*}
e^+\,e^-\,\rightarrow\,q\,\bar{q}\,g,
\end{\eqn*}
virtual corrections, as well as real emission processes
\begin{eqnarray}
&&e^+\,e^-\,\rightarrow\,q\,\bar{q}\,q\,\bar{q},\;e^+\,e^-\,\rightarrow\,q\,\bar{q}\,g\,g.
\end{eqnarray}
The NLO QCD results for this process have first been presented in Refs. \cite{Ellis:1980wv,Kuijf:1991kn,Giele:1993dj,Catani:1996jh}, and the respective expressions for the virtual corrections can be found therein. We want to emphasize that already for this quite simple process, the above real emission contributions call for $(8+10)$ matrix element reevaluations per phase space point in the CS and $(4+5)$ reevaluations in our scheme, respectively\footnote{In fact the number of mappings/subtraction terms can be reduced making use of symmetries; however, this applies to both schemes on the same footing, therefore we still obtain a reduced scaling behavior with respect to the Catani Seymour scheme. We thank M. Seymour for useful comments regarding this.}.
%It is natural to group the results according to their color structure; we then have contributions proportional to
%\begin{\eqn}\label{eq:color}
%N_C\,C_F\,n_f\,T_R,\,N_C^2\,C_F,\,N_C\,C_F^2,
%\end{\eqn} 
%where the $q\bar{q}q\bar{q}$/ $q\bar{q}gg$ final state process contribute to $N_C\,C_F\,n_f\,T_R,\,(N_C\,C_F^2,\,N_C^2\,C_F)$ only. We display our results in terms of the C distribution\cite{Ellis:1980wv} 
We display our results in terms of the C distribution\cite{Ellis:1980wv} 
\begin{\eqn}\label{eq:C}
{\textstyle C^{(n)}\,=\,3\,\left\{ 1-\sum_{i,j\,=\,1,\,i<j}^n\,\frac{s_{ij}^2}{(2\,p_i\cdot\,Q)\,(2\,p_j\cdot\,Q)}  \right\},\,{(s_{ij}\,=\,2\,p_i \cdot p_j)}},
\end{\eqn}
which fulfills all requirements of a jet observable and is infrared finite on the integration level\cite{Catani:1997xc}. 
\begin{figure}
\begin{minipage}{0.49\textwidth}
\begin{center}
\includegraphics[width=\textwidth]{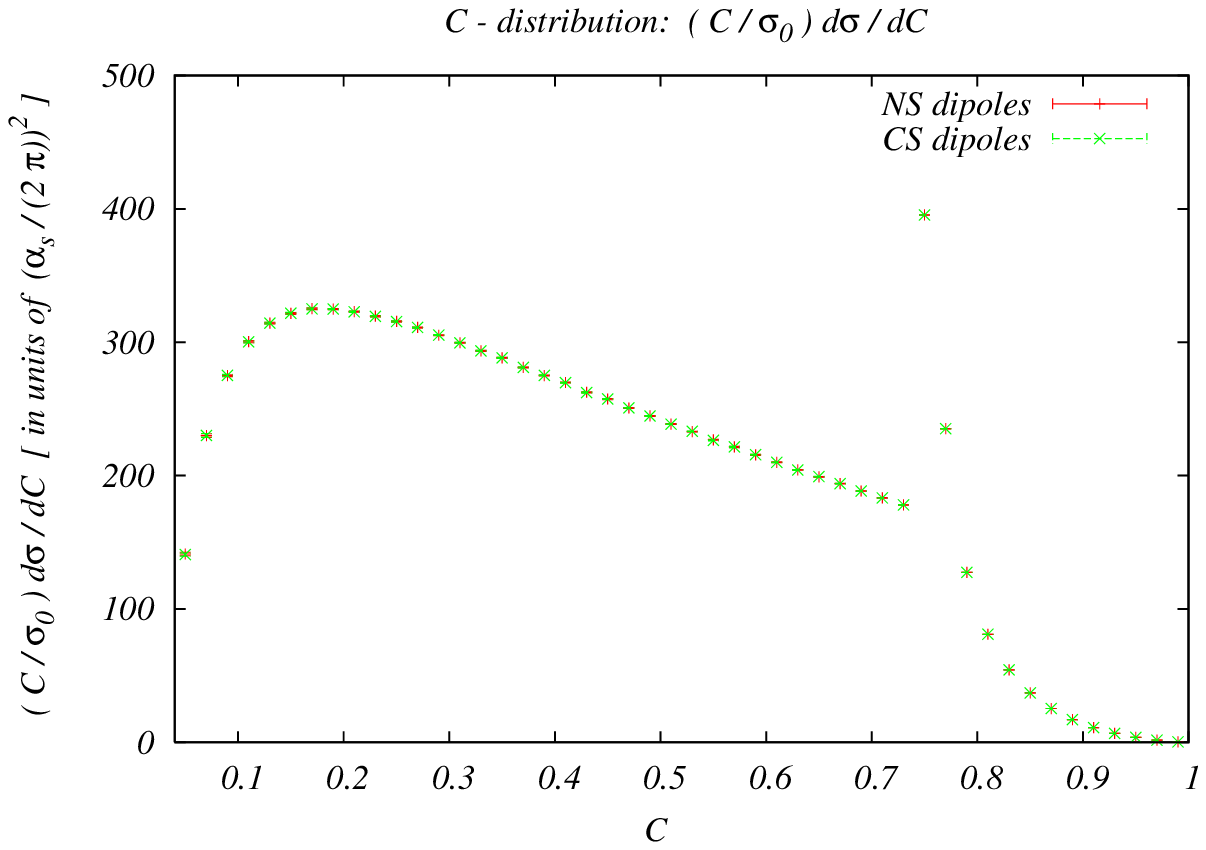}
%\caption{\label{fig:tot} Total result for differential distribution $\frac{C}{\sigma_0}\,\frac{d\sigma^\text{NLO}}{dC}$ using both NS (red) and CS (green) dipoles. The standard literature result obtained using the CS scheme is completely reproduced with the NS dipoles.}
\end{center}
\end{minipage}
%\end{figure}
%\begin{figure}
\begin{minipage}{0.49\textwidth}
\begin{center}
\includegraphics[width=\textwidth]{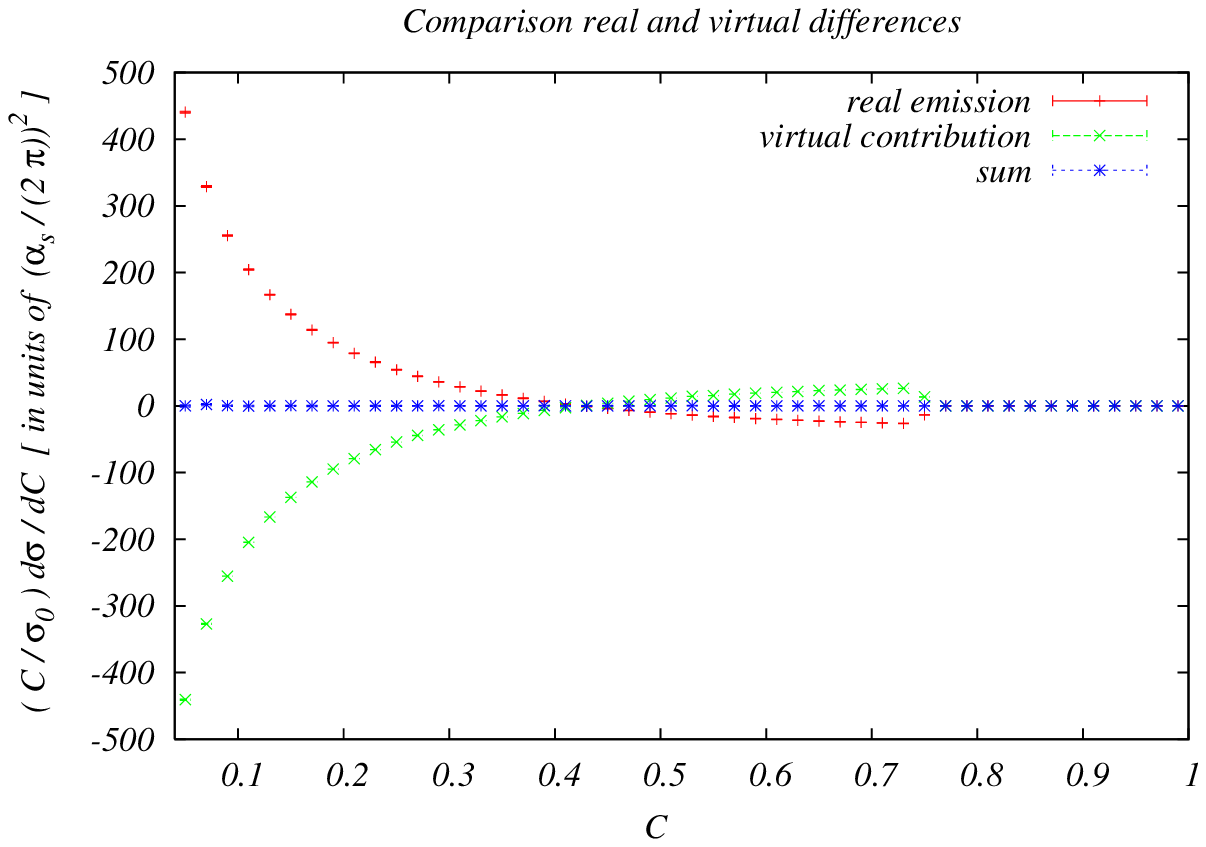}
\end{center}
\end{minipage}
\caption{\label{fig:tot_diff} {\sl Left:} Total result for differential distribution $\frac{C}{\sigma_0}\,\frac{d\sigma^\text{NLO}}{dC}$ using both NS (red) and CS (green) dipoles. The standard literature result obtained using the CS scheme is completely reproduced with the NS dipoles. {\sl Right:} {\sl Differences} $\Delta_\text{CS-NS}$ for real emission (red, upper) and virtual (green, lower) contributions, showing that especially for low $C$ values the contributions in the two schemes significantly differ. Adding up $\Delta^\text{real}+\Delta^\text{virt}$ gives 0 as expected.}
%\end{center}
\end{figure}
Figure \ref{fig:tot_diff} shows that we reproduce the literature result, numerically obtained from Ref. \cite{mike}, as well as agreement between implementations of both schemes (results in terms of distinct color subsets, where the actual per mil-level agreement becomes apparent, are displayed in Fig 3 of Ref. \cite{Chung:2012rq}). We want to point out that this is indeed a non-trivial statement, as the {\sl differences} between the two schemes for both subtracted real emission as well as virtual contributions are sizeable; therefore, agreement between the two schemes on the per mil level as presented in Ref. \cite{Chung:2012rq} constitutes a non-trivial validation of our scheme. 
\subsection{Massive emitters}
For massive emitters, the collinear singularities are generically regulated by the mass of the emitter, and therefore the pole structure is reduced to single poles $\sim\,\vareps^{-1}$ in this case. On the other hand, the parametrization of the integration measure as well as the respective mappings are more involved. In addition, splitting functions are no longer invariant under $q\,\longleftrightarrow\,\bar{q}$, which in principle calls for evaluation of five additional contributions. Details on these features in the case of massive emitters become already apparent in the discussion of the parton shower in Ref. \cite{Nagy:2007ty}, which has been formulated including full mass-dependence. Obviously, all advantages of the massless scheme, especially regarding the improved scaling behavior and the matching to the parton shower, persist. We here want to point to current work undergoing within the Helac-NLO framework \cite{Bevilacqua:2011xh} to include this scheme in a completely automatized way, c.f. Ref. \cite{Bevilacqua:2013taa}. However, the authors have not yet published analytic expressions in the massive scheme; to our understanding, this is still work in progress.

\section{Summary}\label{sec:summ}

In this review, we have discussed the current status of an alternative NLO subtraction scheme for QCD calculations, which uses the splitting functions of an improved parton shower as subtraction kernels. We have revised the setup, and especially the features leading to an improved scaling behavior of our scheme with respect to one of the standard subtraction proposals. We equally pointed to possible further improvements of this scaling behavior, using symmetries of specific processes which have first been largely investigated within the MadFKS framework. We also list reasons why our scheme should in principle account for an improvement in the matching of NLO calculations with a parton shower using splitting functions which correspond to the subtraction kernels. We have given examples of the scheme for a process with massless emitters, namely $e^+\,e^-\,\rightarrow\,3$ jets, and pointed to an automatized implementation which also contains subtraction terms for massive emitters.

Apart from these more general features, we have focussed in slightly more detail on the only feature of the massless scheme which might be considered as a drawback, i.e. the fact that some remainders of the integrated subtraction terms are currently not available in an analytic format. We discussed prospects of evaluating these remaining terms analytically or approximating them by polynomials or two-dimensional grids, where we have pointed to current ongoing work in this direction.

Summarizing, we regard the scheme discussed here as a viable alternative to both CS and FKS subtraction. Our scheme exhibits a smaller number of subtraction terms with respect to CS, and does not call for a reparametrization of the phase space for each emitter/emitted parton pair, as needed in the FKS scheme. While currently not yet automatized in a publicly available code, we believe that this direction is worthwhile to investigate, and hope results on this will appear in the near future.

\section*{Acknowledgments}
I thank the Editors of MPLA for inviting this review. Some of the results discussed in this manuscript are based on work of C.H. Chung and M. Bach, to whom I am grateful for fruitful collaboration. I equally thank them as well as C. Gnendiger and D. St\"ockinger for useful comments regarding the manuscript.

\bibliographystyle{utcaps}
\bibliography{NLO_subtraction}

\end{document}